\documentclass[12pt]{article}

\newcommand{\mybar}[1]{\overline{#1}}

\def\smallromani{\renewcommand{\theenumi}{\roman{enumi}}
\renewcommand{\labelenumi}{(\theenumi)}}

\newcommand{\Proof}{\NI
                    {\bf Proof.}\ }

\usepackage{sgame}
\usepackage{url}
\usepackage{amsmath}

\usepackage{thanks}

\usepackage{color}

\usepackage{amsfonts}

\newcommand{\oldbfe}[1]{\begin{bfseries}\emph{#1}\end{bfseries}}

\newcommand{\ES}{\mbox{$\emptyset$}}

\newcommand{\myra}{\mbox{$\:\rightarrow\:$}}

\newcommand{\A}{\mbox{$\ \wedge\ $}}

\newcommand{\sse}{\mbox{$\:\subseteq\:$}}

\newcommand{\fa}{\mbox{$\forall$}}
\newcommand{\te}{\mbox{$\exists$}}

\newcommand{\LL}{\mbox{$\ldots$}}

\newcommand{\C}[1]{\mbox{$\{{#1}\}$}}           

\newcommand{\NI}{\noindent}
\newcommand{\HB}{\hfill{$\Box$}}
\newcommand{\VV}{\vspace{5 mm}}

\newcommand{\II}{\vspace{2 mm}}

\newcommand{\restr}[1]{\! \mid \! {#1}}

\newcommand{\szkew}[1]{\relax \setbox0=\hbox{\kern -24pt $\displaystyle#1$\kern 0pt }%
\box0}
{\catcode`\@=11 \global\let\ifjusthvtest@=\iffalse}

\newcounter{oldmycaption}

\newtheorem{theorem}{Theorem}
\newtheorem{defined}{Definition}

\newtheorem{exa}{Example}

\newtheorem{lemma}{Lemma}
\newtheorem{corollary}{Corollary}

\newtheorem{note}{Note}

\title{The Role of Monotonicity in the Epistemic Analysis of Strategic Games}

 \author{Krzysztof R. Apt \thanksref{t1}\thanksref{t2} and Jonathan A. Zvesper\thanksref{t3}}
\thankstext{t1}{CWI, Science Park 123, 1098 XG Amsterdam, The Netherlands}
\thankstext{t2}{University of Amsterdam}
\thankstext{t3}{Oxford University, Computing Laboratory, Wolfson
    Building, Parks Road, Oxford OX1 3QD, UK}

\date{}
\begin{document}

\maketitle

\begin{abstract}
  It is well-known that in finite strategic games true common belief
  (or common knowledge) of rationality implies that the players will
  choose only strategies that survive the iterated elimination of
  strictly dominated strategies.  We establish a general theorem that
  deals with monotonic rationality notions and arbitrary strategic
  games and allows to strengthen the above result to arbitrary games,
  other rationality notions, and transfinite iterations of the
  elimination process.  We also clarify what conclusions one can draw
  for the customary dominance notions that are not monotonic. The main
  tool is Tarski's Fixpoint Theorem.
\end{abstract}



\section{Introduction}
\label{sec:intro}

\subsection{Contributions}

In this paper we provide an epistemic analysis of arbitrary strategic
games based on possibility correspondences. We prove a general result
that is concerned with monotonic program properties\footnote{The
  concept of a monotonic property is introduced in
  Section \ref{sec:prelim}.} used by the players to select optimal
strategies.

More specifically, given a belief model for the initial strategic
game, denote by $\textbf{RAT}(\phi)$ the property that
each player $i$ uses a property $\phi_i$ to select his strategy (`each
player $i$ is $\phi_i$-rational').
We establish in Section \ref{sec:epistemic} the following general result:

  Assume that each property $\phi_i$ is monotonic.  The
  set of joint strategies that the players choose in the states in which
  $\textbf{RAT}(\phi)$ is a true common belief is included
  in the set of joint strategies that remain after the iterated
  elimination of the strategies that for player $i$ are not
  $\phi_i$-optimal.

In general, transfinite iterations of the strategy elimination are possible.
For some belief models the inclusion can be reversed.

This general result covers the usual notion of rationalizability in
finite games and a `global' version of the iterated elimination of
strictly dominated strategies used in \cite{MR90} and studied for
arbitrary games in \cite{CLL07}.  It does not hold for the `global'
version of the iterated elimination of weakly dominated strategies.
For the customary, `local' version of the iterated elimination of
strictly dominated strategies we justify in Section
\ref{sec:consequences} the statement
\begin{quote}
  \emph{true common belief (or common knowledge) of rationality implies that
  the players will choose only strategies that survive the iterated
  elimination of strictly dominated strategies}
\end{quote}
for arbitrary games and transfinite iterations of the elimination
process. Rationality refers here to the concept studied in
\cite{Ber84}.  We also show that the above general result yields a
simple proof of the well-known version of the above result for finite
games and strict dominance by a mixed strategy.

The customary, local, version of strict dominance is non-monotonic, so
the use of monotonic properties has allowed us to provide epistemic
foundations for a non-monotonic property.  However, weak dominance,
another non-monotonic property, remains beyond the reach of this
approach.  In fact, we show that in the above statement we cannot
replace strict dominance by weak dominance.  A mathematical reason is
that its global version is also non-monotonic, in contrast to strict
dominance, the global version of which is monotonic.  To provide
epistemic foundations of weak dominance the only currently known
approaches are \cite{BFK08} based on lexicographic probability
systems and \cite{HP09} based on a version of the `all I know'
modality.


\subsection{Connections}

The relevance of monotonicity in the context of epistemic analysis of
finite strategic games has already been pointed out in \cite{vB07}.
The distinction between local and global properties is from \cite{Apt07}
and \cite{Apt07c}.

To show that for some belief models an equality holds between the set
of joint strategies chosen in the states in which
$\textbf{RAT}({\phi})$ is true common belief and the set of joint
strategies that remain after the iterated elimination of the
strategies that for player $i$ are not $\phi_i$-rational requires use
of transfinite ordinals.  This complements the findings of
\cite{Lip91} in which transfinite ordinals are used in a study of
limited rationality, and \cite{Lip94}, where a two-player game is
constructed for which the $\omega_0$ (the first infinite ordinal) and
$\omega_{0} + 1$ iterations of the rationalizability operator of
\cite{Ber84} differ.

In turn, \cite{HS98} show that arbitrary
ordinals are necessary in the epistemic analysis of arbitrary
strategic games based on partition spaces.
Further, as shown in \cite{CLL07}, the global version of the iterated
elimination of strictly dominated strategies, when used for arbitrary
games, also requires transfinite iterations of the underlying
operator.

Finally, \cite{Luo01} invokes Tarski's Fixpoint Theorem, in the
context of what the author calls ``general systems'', and uses this to
prove that the set of rationalizable strategies in a finite
non-cooperative game is the largest fixpoint of a certain operator.
That operator coincides with the global version of the elimination of
never-best-responses.

Some of the results presented here were initially reported in a
different presentation, in \cite{Apt07a}.

\section{Preliminaries}
\label{sec:prelim}


\subsection{Strategic Games}

Given $n$ players ($n > 1$) by a \oldbfe{strategic game} (in short, a
\oldbfe{game}) we mean a sequence 
$
(S_1, \LL, S_n, p_1, \LL, p_n),
$
where for all $i \in \{1, \LL, n\}$

\begin{itemize}
\item $S_i$ is the non-empty set of \oldbfe{strategies}
available to player $i$,

\item $p_i$ is the \oldbfe{payoff function} for the  player $i$, so
$
p_i : S_1 \times \LL \times S_n \myra \cal{R},
$
where $\cal{R}$ is the set of real numbers.
\end{itemize}

We denote the strategies of player $i$ by $s_i$, possibly with some
superscripts.  We call the elements of $S_1 \times \LL \times S_n$ \oldbfe{joint strategies}.
Given a joint strategy $s$ we denote the
$i$th element of $s$ by $s_i$, write sometimes $s$ as $(s_i, s_{-i})$,
and use the following standard notation:

\begin{itemize}
\item $s_{-i} := (s_1, \LL, s_{i-1}, s_{i+1}, \LL, s_n)$,

\item $S_{-i} := S_1 \times \LL \times S_{i-1} \times S_{i+1} \times \LL \times S_n$.

\end{itemize}

Given a finite non-empty set $A$ we denote by
$\Delta A$ the set of probability distributions over $A$ and call
any element of $\Delta S_i$ a \oldbfe{mixed strategy} of player $i$.

In the remainder of the paper we assume an initial strategic game
\[
H := (H_1, \LL, H_n, p_1, \LL, p_n)
\]
A \oldbfe{restriction} of $H$ is a sequence $(G_1, \LL, G_n)$ such that
$G_i \sse H_i$  for all $i \in \{1, \LL, n\}$.
Some of $G_i$s can be the empty set.
We identify the restriction  $(H_1, \LL, H_n)$ with $H$.
We shall focus on the complete lattice
that consists of the set of all restrictions of the game $H$
ordered by the componentwise set inclusion:
\[
\mbox{$(G_1, \LL, G_n) \sse (G'_1, \LL, G'_n)$ iff $G_i \sse G'_i$ for all $i \in \{1, \LL, n\}$}
\]
So in this lattice $H$ is the largest element in this lattice.

\subsection{Possibility Correspondences}
\label{subsec:poss}

In this and the next subsection we essentially follow the survey of \cite{BB99}.
Fix a non-empty set $\Omega$ of \oldbfe{states}.
By an \oldbfe{event} we mean a subset of $\Omega$.

A \oldbfe{possibility correspondence} is a mapping from $\Omega$ to
the powerset ${\cal P}(\Omega)$ of $\Omega$.
We consider three properties of a possibility correspondence $P$:

\begin{enumerate}\smallromani
\item for all $\omega$, $P(\omega) \neq \ES$,

\item for all $\omega$ and $\omega'$, $\omega' \in P(\omega)$ implies $P(\omega') = P(\omega)$,

\item for all $\omega$, $\omega \in P(\omega)$.
\end{enumerate}

If the possibility correspondence satisfies properties (i) and (ii),
we call it a \oldbfe{belief correspondence} and if it satisfies
properties (i)--(iii), we call it a \oldbfe{knowledge
  correspondence}.\footnote{Note that the notion of a belief has two
  meanings in the literature on epistemic analysis of strategic games,
  so also in this paper.  From the context it is always clear which
  notion is used.  In the modal logic terminology a belief
  correspondence is a frame for the modal logic KD45 and a knowledge
  correspondence is a frame for the modal logic S5, see, e.g.
  \cite{BRV01}.}  Note that each knowledge correspondence $P$ yields a
partition $\{P(\omega) \mid \omega \in \Omega\}$ of $\Omega$.

Assume now that each player $i$ has at its disposal a possibility
correspondence $P_i$.
Fix an event $E$. We define
\[
\square E := \square^1 E := \{\omega \in \Omega \mid
\fa i \in \{1, \LL, n\} \: P_i(\omega) \sse E\}
\]
by induction on $k \geq 1$
\[
\square^{k+1} E := \square \square^{k} E
\]
and finally
\[
\square^* E := \bigcap_{k =1}^{\infty} \square^{k} E
\]

If all $P_i$s are belief correspondences, we usually write $B$ instead
of $\square$ and if all $P_i$s are knowledge correspondences, we
usually write $K$ instead of $\square$.  When $\omega \in B^* E$, we
say that the event $E$ is \oldbfe{common belief in the state $\omega$}
and when $\omega \in K^* E$, we say that the event $E$ is
\oldbfe{common knowledge in the state $\omega$}.


An event $F$ is called \oldbfe{evident} if $F \sse \square F$.  That
is, $F$ is evident if for all $\omega \in F$ we have $P_i(\omega) \sse
F$ for all $i \in \{1, \LL, n\}$.  In what follows we shall use the
following alternative characterizations of common belief and common
knowledge based on evident events:
\begin{equation}
  \begin{array}{l}
\mbox{$\omega \in \square^* E$ iff for some evident event $F$ we have $\omega \in F \sse \square E$}
  \end{array}
\label{equ:cb}
\end{equation}
where $\square = B$ or $\square = K$ (see \cite{MS89}, respectively
Proposition 4 on page 180 and Proposition on page 174), and
\begin{equation}
  \label{equ:ck}
\mbox{$\omega \in K^* E$ iff for some evident event $F$ we have $\omega \in F \sse E$}
\end{equation}
(\cite{Aum76}, page 1237).

\subsection{Models for Games}

We now relate these considerations to strategic games.  Given a
restriction $G := (G_1, \LL, G_n)$ of the initial game $H$, by a
\oldbfe{model} for $G$ we mean a set of states $\Omega$ together with
a sequence of functions $\mybar{s_i}: \Omega \myra G_i$, where $i \in
\{1, \LL, n\}$. We denote it by $(\Omega, \mybar{s_1}, \LL, \mybar{s_n})$.

In what follows, given a function $f$ and a subset $E$ of
its domain, we denote by $f(E)$ the range of $f$ on $E$ and by $f
\restr{E}$ the restriction of $f$ to $E$.

By the \oldbfe{standard model} ${\cal M}$ for $G$ we mean the model in which

\begin{itemize}
 \item $\Omega := G_1 \times \LL \times G_n$

 \item $\mybar{s_i}(\omega) := \omega_i$, where $\omega = (\omega_1, \LL, \omega_n)$
\end{itemize}
So the states of the standard model for $G$ are exactly the joint strategies in $G$,
and each $\mybar{s_i}$ is a projection function.
Since the initial game $H$ is given, we know the payoff functions $p_1,
\LL, p_n$. So in the context of $H$ the standard model is an alternative way of
representing a restriction of $H$.

Given a (not necessarily standard) model ${\cal M} := (\Omega, \mybar{s_1}, \LL, \mybar{s_n})$ for a restriction
$G$ and a sequence of events $\overline{E} = (E_1, \LL, E_n)$ in ${\cal
  M}$ (\emph{i.e.}, of subsets of $\Omega$) we define
\[
G_{\overline{E}} := (\mybar{s_1}(E_1), \LL, \mybar{s_n}(E_n))
\]
and call it the \oldbfe{restriction of $G$ to $\overline{E}$}.
When each $E_i$ equals $E$ we write $G_{E}$ instead of $G_{\overline{E}}$.

Finally, we extend the notion of a model for a restriction $G$ to a
\oldbfe{belief model} for $G$ by assuming that each player $i$ has a
belief correspondence $P_i$ on $\Omega$. If each $P_i$ is a knowledge
correspondence, we refer then to a \oldbfe{knowledge model}.
We write each belief model as
\[
(\Omega, \mybar{s_1}, \LL, \mybar{s_n}, P_1, \LL, P_n)
\]

\subsection{Operators}

Consider a fixed complete lattice $(D, \sse)$ with the largest element $\top$.
In what follows we use ordinals and denote them by $\alpha, \beta, \gamma$.
Given a, possibly transfinite, sequence $(G_{\alpha})_{\alpha < \gamma}$ of
elements of $D$ we denote their join and meet respectively by
$\bigcup_{\alpha < \gamma} G_{\alpha}$
and $\bigcap_{\alpha < \gamma} G_{\alpha}$.

Let $T$ be an operator on $(D, \sse)$, \emph{i.e.}, $T: D \myra D$.

\begin{itemize}

\item We call $T$
\oldbfe{monotonic} if for all $G, G'$,
$G \sse G'$ implies $T(G) \sse T(G')$, and \oldbfe{contracting} if for all $G$,
$T(G) \sse G$.

\item We say that an element $G$ is a \oldbfe{fixpoint} of $T$ if $G = T(G)$
and a \oldbfe{post-fixpoint} of $T$ if $G \sse T(G)$.

\item We define by
transfinite induction a sequence of elements $T^{\alpha}$ of $D$, where $\alpha$ is an ordinal, as follows:

\begin{itemize}

  \item $T^{0} := \top$,

  \item $T^{\alpha+1} := T(T^{\alpha})$,

  \item for all limit ordinals $\beta$, $T^{\beta} := \bigcap_{\alpha < \beta} T^{\alpha}$.
  \end{itemize}

\item We call the least $\alpha$ such that $T^{\alpha+1} = T^{\alpha}$ the \oldbfe{closure ordinal} of $T$
and denote it by $\alpha_T$.  We call then $T^{\alpha_T}$ the \oldbfe{outcome of} (iterating) $T$ and write it alternatively as $T^{\infty}$.
\end{itemize}

So an outcome is a fixpoint reached by a transfinite iteration that
starts with the largest element.  In general, the outcome of an
operator does not need to exist but we have the following classic
result due to \cite{Tar55}.\footnote{We use here its `dual' version in
  which the iterations start at the largest and not at the least
  element of a complete lattice.}
\II

\NI
\textbf{Tarski's Fixpoint Theorem}
Every monotonic operator $T$ on $(D, \sse)$
has an outcome, \emph{i.e.}, $T^{\infty}$ is well-defined.
Moreover,
\[
T^{\infty} = \nu T = \cup \{G \mid G \sse T(G)\}
\]
where $\nu T$ is the largest fixpoint of $T$.
\vspace{2mm}

In contrast, a contracting operator does not need to have a largest fixpoint.
But we have the following obvious observation.

\begin{note} \label{note:con}
Every contracting operator $T$ on $(D, \sse)$ has an outcome, \emph{i.e.},
$T^{\infty}$ is well-defined.
\HB
\end{note}

In Section \ref{sec:consequences} we shall need the following lemma, that modifies the corresponding
lemma from \cite{Apt07c} from finite to arbitrary complete lattices.

\begin{lemma} \label{lem:inc}
Consider two operators $T_1$ and $T_2$ on $(D, \sse)$ such that
\begin{itemize}
\item for all $G$, $T_1(G) \sse T_2(G)$,

\item $T_1$ is monotonic,

\item $T_2$ is contracting.
\end{itemize}
Then $T_1^{\infty} \sse T_2^{\infty}$.
\end{lemma}
\Proof
We first prove by transfinite induction that for all $\alpha$
\begin{equation}
T_1^{\alpha} \sse T_2^{\alpha}
  \label{equ:inc}
\end{equation}

By the definition of the iterations we only need to consider the induction
step for a successor ordinal.  So suppose the claim holds for some
$\alpha$. Then by the first two assumptions and the induction
hypothesis we have the following string of inclusions and equalities:
\[
T_1^{\alpha + 1} =   T_1(T_1^{\alpha}) \sse T_1(T_2^{\alpha}) \sse T_2(T_2^{\alpha}) = T_2^{\alpha + 1}
\]

This shows that for all $\alpha$ (\ref{equ:inc}) holds.
By Tarski's Fixpoint Theorem and Note \ref{note:con} the outcomes of
$T_1$ and $T_2$ exist, which implies the claim.
\HB

\subsection{Iterated Elimination of Non-Rational Strategies}
\label{subsec:iter}

In this paper we are interested in analyzing situations in which each
player pursues his own notion of rationality and
this information is common knowledge or true common
belief.  As a special case we cover then the usually analyzed
situation in which all players use the same notion of rationality.

Given player $i$ in the initial strategic game $H := (H_1, \LL, H_n, p_1,
\LL,p_n)$ we formalize his notion of rationality using an \oldbfe{optimality property}
$\phi(s_i, G_i, G_{-i})$ that holds between a strategy $s_i \in H_i$, a set $G_i$ of strategies
of player $i$ and a set $G_{-i}$ of joint strategies of his opponents.
Intuitively, $\phi_{i}(s_i, G_i, G_{-i})$
holds if $s_i$ is an `optimal' strategy for player $i$ within the
restriction $G := (G_i, G_{-i})$, assuming that he uses the property
$\phi_i$ to select optimal strategies. In Section \ref{sec:consequences}
we shall provide several natural examples of such properties.

We say that the property $\phi_{i}$ used by player $i$
is \oldbfe{monotonic} if
for all $G_{-i}, G'_{-i} \sse H_{-i}$ and $s_i \in H_i$
\[
\mbox{$G_{-i} \sse G'_{-i}$ and $\phi(s_i, H_i, G_{-i})$ imply $\phi(s_i, H_i, G'_{-i})$}
\]
So monotonicity refers to the situation in which the set of strategies of player $i$ is set to $H_i$ and
the set of joint strategies of player $i$'s opponents is increased.

Each sequence of properties $\phi := (\phi_1, \LL, \phi_n)$
determines an operator $T_{\phi}$ on the restrictions of
$H$ defined by
\[
T_{\phi}(G) := G'
\]
where $G := (G_1, \LL, G_n)$, $G' := (G'_1, \LL, G'_n)$, and for all $i \in \{1, \LL, n\}$
\[
G'_i := \{ s_i \in G_i \mid \phi_i(s_i, H_i, G_{-i})\}
\]

Note that in defining the set of strategies $G'_i$ we use in the
second argument of $\phi_i$ the set $H_i$ of player's $i$ strategies
in the \emph{initial} game $H$ and not in the \emph{current}
restriction $G$. This captures the idea that at every stage of the
elimination process player $i$ analyzes the status of each strategy in
the context of his initial set of strategies.

Since $T_{\phi}$ is contracting, by Note \ref{note:con} it
has an outcome, \emph{i.e.}, $T_{\phi}^{\infty}$ is well-defined.
Moreover, if each $\phi_i$ is monotonic, then $T_{\phi}$ is
monotonic and by Tarski's Fixpoint Theorem its largest fixpoint $\nu
T_{\phi}$ exists and equals $T_{\phi}^{\infty}$.
Finally, $G$ is a fixpoint of $T_{\phi}$ iff for all
$i \in \{1, \LL, n\}$ and all $s_i \in G_i$, $\phi_i(s_i, H_i,
G_{-i})$ holds.

Intuitively, $T_{\phi}(G)$ is the result of removing from
$G$ all strategies that are not $\phi_i$-rational. So the outcome of
$T_{\phi}$ is the result of the iterated elimination of
strategies that for player $i$ are not $\phi_i$-rational.

\section{Two Theorems}
\label{sec:epistemic}

We now assume that each player $i$ employs some
property $\phi_i$ to select his strategies, and we analyze the
situation in which this information is true common belief or common
knowledge.  To determine which strategies are then selected by the
players we shall use the $T_{\phi}$ operator.

We begin by fixing a belief model $(\Omega, \mybar{s_1}, \LL,
\mybar{s_n}, P_1, \LL, P_n)$ for the initial game $H$.  Given an
optimality property $\phi_i$ of player $i$ we say that player $i$ is
$\phi_i$-\oldbfe{rational in the state} $\omega$ if
$\phi_i(\mybar{s_{i}}(\omega), H_i, (G_{P_i(\omega)})_{-i})$ holds.
Note that when player $i$ believes (respectively, knows) that the
state is in $P_i(\omega)$, the set $(G_{P_i(\omega)})_{-i}$ represents
his belief (respectively, his knowledge) about other players'
strategies.  That is, $(H_i, (G_{P_i(\omega)})_{-i})$ is the
restriction he believes (respectively, knows) to be relevant to his
choice.

Hence $\phi_i(\mybar{s_{i}}(\omega), H_i, (G_{P_i(\omega)})_{-i})$
captures the idea that if player $i$ uses $\phi_i$ to select his
strategy in the game he considers relevant, then in the state $\omega$
he indeed acts `rationally'.

To reason about common knowledge and true common belief we introduce the
event
\[
\mbox{$\textbf{RAT}({\phi}) := \{\omega \in \Omega \mid $ each player $i$ is $\phi_i$-rational in $\omega$\}}
\]
and consider the following two events constructed out of it:
$K^* \textbf{RAT}({\phi})$ and
$\textbf{RAT}({\phi}) \cap B^* \textbf{RAT}({\phi})$.
We then focus on the corresponding restrictions $G_{K^* \textbf{RAT}({\phi})}$
and $G_{\textbf{RAT}({\phi}) \cap B^* \textbf{RAT}({\phi})}$.

So strategy $s_i$ is an element of the $i$th component of
$G_{K^* \textbf{RAT}({\phi})}$ if
$s_i = \mybar{s_i}(\omega)$ for some
$\omega \in K^* \textbf{RAT}({\phi})$. That is,
$s_i$ is a strategy that player $i$ chooses in a state in which
it is common knowledge that each player $j$ is $\phi_j$-rational, and similarly for
$G_{\textbf{RAT}({\phi}) \cap B^* \textbf{RAT}({\phi})}$.

The following result then relates for arbitrary strategic games the
restrictions $G_{\textbf{RAT}({\phi}) \cap B^*
  \textbf{RAT}({\phi})}$ and $G_{K^*
  \textbf{RAT}({\phi})}$ to the outcome of the iteration of
the operator $T_{\phi}$.

\begin{theorem} \label{thm:epist1}
\mbox{}\\[-6mm]
\begin{enumerate} \smallromani
\item Suppose that each property $\phi_i$ is monotonic. Then for all belief models for $H$
\[
G_{\textbf{RAT}({\phi}) \cap B^* \textbf{RAT}({\phi})} \sse T_{\phi}^{\infty}
\]

\item Suppose that each property $\phi_i$ is monotonic. Then for all knowledge models for $H$
\[
G_{K^* \textbf{RAT}({\phi})} \sse T_{\phi}^{\infty}
\]

\item For some standard knowledge model for $H$
\[
T_{\phi}^{\infty} \sse G_{K^* \textbf{RAT}({\phi})}
\]
\end{enumerate}
\end{theorem}

So part $(i)$ (respectively, $(ii)$) states that true common
belief (respectively, common knowledge) of $\phi_i$-rationality of
each player $i$ implies that the players will choose only strategies
that survive the iterated elimination of non-$\phi$-rational
strategies.  \II

\newpage

\Proof

\NI
$(i)$ Fix a belief model
$(\Omega, \mybar{s_1}, \LL, \mybar{s_n}, P_1, \LL, P_n)$ for $H$.
Take a strategy $s_i$ that is an element of the $i$th component of
$G_{\textbf{RAT}({\phi}) \cap B^* \textbf{RAT}({\phi})}$.
Thus we have $s_i = \mybar{s_i}(\omega)$ for some state $\omega$ such that
$\omega \in \textbf{RAT}({\phi})$ and $\omega \in B^* \textbf{RAT}({\phi})$.
The latter implies by (\ref{equ:cb}) that for some evident event $F$
\begin{equation}
  \label{equ:F}
\omega \in F \sse \{\omega' \in \Omega \mid \fa i \in \{1, \LL, n\} \: P_i(\omega') \sse \textbf{RAT}({\phi})\}
\end{equation}

Take now an arbitrary $\omega' \in F \cap
\textbf{RAT}({\phi})$ and $i \in \{1, \LL, n\}$.  Since
$\omega' \in \textbf{RAT}({\phi})$, it holds that player $i$ is
$\phi_i$-rational in $\omega'$, \emph{i.e.}, $\phi_i(\mybar{s_i}(\omega'), H_i,
(G_{P_i(\omega')})_{-i})$ holds.  But $F$ is evident, so $P_i(\omega') \sse
F$. Moreover by (\ref{equ:F}) $P_i(\omega') \sse
\textbf{RAT}({\phi})$, so $P_i(\omega') \sse F \cap
\textbf{RAT}({\phi})$.  Hence $(G_{P_i(\omega')})_{-i} \sse
(G_{F \cap \textbf{RAT}({\phi})})_{-i}$ and by the monotonicity of
$\phi_i$ we conclude that $\phi_i(\mybar{s_i}(\omega'), H_i, (G_{F \cap
  \textbf{RAT}({\phi})})_{-i})$ holds.

By the definition of $T_{\phi}$ this means that $G_{F \cap
  \textbf{RAT}({\phi})} \sse T_{\phi}(G_{F \cap
  \textbf{RAT}({\phi})})$, \emph{i.e.}  $G_{F \cap \textbf{RAT}({\phi})}$ is
a post-fixpoint of $T_{\phi}$.  But $T_{\phi}$ is monotonic since each
property $\phi_i$ is.  Hence by Tarski's Fixpoint Theorem $G_{F \cap
  \textbf{RAT}({\phi})} \sse T_{\phi}^{\infty}$.  But $s_i =
\mybar{s_i}(\omega)$ and $\omega \in F \cap {\textbf{RAT}({\phi})}$,
so we conclude by the above inclusion that $s_i$ is an element of the
$i$th component of $T_{\phi}^{\infty}$.  This proves the claim.  \II

\NI
$(ii)$
By the definition of common knowledge for all events $E$ we have $K^* E \sse E$. Hence
for all $\phi$ we have
$K^* \textbf{RAT}({\phi}) \sse \textbf{RAT}({\phi}) \cap K^* \textbf{RAT}({\phi})$
and consequently
$G_{K^* \textbf{RAT}({\phi})} \sse G_{\textbf{RAT}({\phi}) \cap K^* \textbf{RAT}({\phi})}$.

So part (ii) follows from part (i).
\II

\NI
$(iii)$
Suppose $T^{\infty}_{\phi} = (G_1, \LL, G_n)$. Consider the event
$F := G_1 \times \LL \times G_n$ in the standard model for $H$. Then
$G_F = T^{\infty}_{\phi}$.
Define each possibility correspondence $P_i$ by
\[
        P_i(\omega) :=
        \left\{
        \begin{array}{l@{\extracolsep{3mm}}l}
        F    & \mathrm{if}\  \omega \in F \\
        \Omega \setminus F       & \mathrm{otherwise}
        \end{array}
        \right.
\]
Each $P_i$ is a knowledge correspondence (also when $F = \ES$ or $F = \Omega$)
and clearly $F$ is an evident event.

Take now an arbitrary $i \in \{1, \LL, n\}$ and an arbitrary state
$\omega \in F$.  Since $T^{\infty}_{\phi}$ is a fixpoint of $T_{\phi}$
and $\mybar{s_i}(\omega) \in G_i$ we have
$\phi_{i}(\mybar{s_i}(\omega), H_i, (T^{\infty}_{\phi})_{-i})$, so by
the definition of $P_{i}$ we have $\phi_{i}(\mybar{s_{i}}(\omega),
H_i, (G_{P_{i}(\omega)})_{-i})$.  This shows that each player $i$ is
$\phi_i$-rational in each state $\omega \in F$, \emph{i.e.}, $F \sse
\textbf{RAT}(\phi)$.

Since $F$ is evident, we conclude by (\ref{equ:ck}) that in each state $\omega \in F$
it is common knowledge that each player $i$ is $\phi_i$-rational, \emph{i.e.},
$F \sse K^* \textbf{RAT}(\phi)$.
Consequently
\[
T_{\phi}^{\infty} = G_F \sse G_{K^*  \textbf{RAT}(\phi)}
\]
\HB
\VV

Items $(i)$ and $(ii)$ show that when each property $\phi_i$ is
monotonic, for all belief models of $H$ it holds that the joint
strategies that the players choose in the states in which each player
$i$ is $\phi_i$-rational and it is common belief that each player $i$
is $\phi_i$-rational (or in which it is common knowledge that each
player $i$ is $\phi_i$-rational) are included in those that remain
after the iterated elimination of the strategies that are not
$\phi_i$-rational.

Note that monotonicity of the $\phi_i$ properties was not needed to
establish item $(iii)$.

By instantiating the $\phi_i$'s with specific properties we get instances of
the above result that refer to specific definitions of rationality.
This will allow us to relate the above result to the ones established
in the literature. Before we do this we establish a result that
identifies a large class of properties $\phi_i$ for which Theorem \ref{thm:epist1}
does not apply.







\begin{theorem} \label{thm:epist}
Suppose that a joint strategy $s \not\in T_{\phi}^{\infty}$ exists such that
\[
\phi_i(s_i, H_i, (\{s_j\}_{j \neq i}))
\]
holds all $i \in \{1, \LL, n\}$.
Then for some knowledge model for $H$ the inclusion
\[
G_{K^* \textbf{RAT}({\phi})} \sse T_{\phi}^{\infty}
\]
does not hold.
\end{theorem}

\Proof
We extend the
standard model for $H$ by the knowledge correspondences $P_1, \LL,
P_n$ where for all $i \in \{1, \LL, n\}$, $P_i(\omega) = \C{\omega}$.
Then for all $\omega$ and all $i \in \{1, \LL, n\}$
\[
G_{P_i(\omega)} = (\C{\mybar{s_1}(\omega)}, \LL, \C{\mybar{s_n}(\omega)})
\]
Let $\omega' := s$. Then for all $i \in \{1, \LL, n\}$,
$G_{P_i(\omega')} = (\C{s_1}, \LL, \C{s_n})$, so
by the assumption each player $i$ is $\phi_i$-rational in $\omega'$, \emph{i.e.},
$\omega' \in \textbf{RAT}(\phi)$. By the definition of
$P_i$s the event $\C{\omega'}$ is evident and $\omega' \in K \textbf{RAT}(\phi)$.
So by (\ref{equ:cb})
$\omega' \in K^* \textbf{RAT}(\phi)$.
Consequently $s = (\mybar{s_1}(\omega'), \LL, \mybar{s_n}(\omega')) \in G_{K^* \textbf{RAT}(\phi)}$.

This yields the desired conclusion by the choice of $s$.
\HB

\section{Applications}
\label{sec:consequences}

We now analyze to what customary game-theoretic properties the above two results apply.
By a \oldbfe{belief} of player $i$ about the strategies his opponents play
given the set $G_{-i}$ of their joint strategies
we mean one of the following possibilities:

\begin{itemize}
\item a joint strategy of the opponents of player $i$, \emph{i.e.}, $s_{-i} \in G_{-i}$, called a \oldbfe{point belief},

\item or, in the case the game is finite, a joint mixed strategy of
  the opponents of player $i$ (\emph{i.e.}, $(m_1, \LL, m_{i-1}, m_{i+1},
  \LL, m_n)$, where $m_j \in \Delta G_j$ for all $j \neq i$), called an \oldbfe{independent belief},

\item or, in the case the game is finite, an element of $\Delta G_{-i}$, called
a \oldbfe{correlated belief}.
\end{itemize}

In the second and third case the payoff function $p_i$ can be lifted
in the standard way to an \oldbfe{expected payoff} function $p_i : H_i
\times {\cal B}_{i}(G_{-i}) \myra \cal{R}$, where ${\cal
  B}_{i}(G_{-i})$ is the corresponding set of beliefs of player $i$
held given $G_{-i}$.

We use below the following abbreviations, where $s_i, s'_i \in H_i$
and $G_{-i}$ is a set of the strategies of the opponents of player $i$:

\begin{itemize}
\item (\oldbfe{strict dominance}) $s'_i \succ_{G_{-i}} s_i$ for

$\fa s_{-i} \in G_{-i} \: p_{i}(s'_i, s_{-i}) > p_{i}(s_i, s_{-i})$

\item (\oldbfe{weak dominance}) $s'_i \succ^{w}_{G_{-i}} s_i$ for

$\fa s_{-i} \in G_{-i} \: p_{i}(s'_i, s_{-i}) \geq p_{i}(s_i, s_{-i}) \A \te s_{-i} \in G_{-i} \: p_{i}(s'_i, s_{-i}) > p_{i}(s_i, s_{-i})$

\end{itemize}

In the case of finite games the relations $\succ_{G_{-i}}$ and
$\succ^{w}_{G_{-i}}$ between a mixed strategy and a pure strategy are
defined in the same way.

We now introduce natural examples of the optimality notion.

\begin{itemize}

\item $sd_{i}(s_i, G_i, G_{-i}) \equiv \neg \te s'_i  \in G_i \: s'_i \succ_{G_{-i}} s_i$

\item (assuming $H$ is finite)
$msd_{i}(s_i, G_i, G_{-i}) \equiv \neg \te m'_i \in \Delta G_i \: m'_i
  \succ_{G_{-i}} s_i$

\item $wd_{i}(s_i, G_i, G_{-i}) \equiv \neg \te s'_i \in G_i \: s'_i \succ^{w}_{G_{-i}} s_i$

\item (assuming $H$ is finite)
$mwd_{i}(s_i, G_i, G_{-i}) \equiv \neg \te m'_i \in \Delta G_i \: m'_i \succ^{w}_{G_{-i}} s_i$

\item $br_{i}(s_i, G_i, G_{-i}) \equiv \te \mu_i \in {\cal B}_i(G_{-i}) \:
  \fa s'_i \in G_i \: p_i(s_i, \mu_i) \geq p_i(s'_i, \mu_i)$
\end{itemize}

So $sd_{i}$ and $wd_{i}$ are the customary
notions of strict and weak dominance and $msd_{i}$ and $mwd_{i}$ are
their counterparts for the case of dominance by a mixed strategy.
Note that the notion $br_{i}$ of best response, comes in three
`flavours' depending on the choice of the set ${\cal B}_i(G_{-i})$ of
beliefs.

Consider now the iterated elimination of strategies as
defined in Subsection \ref{subsec:iter}, so \emph{with} the repeated
reference by player $i$ to the strategy set $H_i$.  For the optimality
notion $sd_i$ such a version of iterated elimination was studied in
\cite{CLL07}, for $mwd_i$ it was used in \cite{BFK08}, while for $br_{i}$
it corresponds to the rationalizability notion of \cite{Ber84}.

In \cite{Lip94}, \cite{CLL07} and \cite{Apt07} examples are provided
showing that for the properties $sd_i$ and $br_i$ in general
transfinite iterations (\emph{i.e.}, iterations beyond $\omega_0$) of the
corresponding operator are necessary to reach the outcome.  So to
establish for them part $(iii)$ of Theorem \ref{thm:epist1}
transfinite iterations of the $T_{\phi}$ operator are
necessary.

The following lemma holds.

\begin{lemma} \label{lem:mono}
The properties
$sd_{i}, \ msd_{i}$ and $br_{i}$
are monotonic.
\end{lemma}
\Proof
Straightforward.
\HB
\VV

So Theorem \ref{thm:epist1} applies to the above three properties.  In
contrast, Theorem \ref{thm:epist1} does not apply to the remaining two
properties $wd_{i}$ and $mwd_{i}$, since, as
indicated in \cite{Apt07c}, the corresponding operators
$T_{wd}$ and $T_{mwd}$ are not
monotonic, and hence the properties $wd_{i}$ and $mwd_{i}$ are not monotonic.

In fact, the desired inclusion does not hold and Theorem
\ref{thm:epist} applies to these two optimality properties. Indeed,
consider the following game:

\begin{center}
\begin{game}{2}{2}
       & $L$    & $R$\\
$U$   &$1,1$   &$0,1$\\
$D$   &$1,0$   &$1,1$
\end{game}
\end{center}

Then the outcome of iterated elimination for both $wd_{i}$ and $mwd_{i}$ yields
$G := (\C{D},\C{R})$. Further, we have $wd_1(U, \{U,D\}, \{L\})$ and
$wd_2(L, \{L,R\}, \{U\})$, and analogously for $mwd_1$ and $mwd_2$.

So the joint strategy $(U,L)$ satisfies the conditions of
Theorem \ref{thm:epist} for both $wd_{i}$ and $mwd_{i}$.
Note that this game also furnishes an example for non-monotonicity of $wd_i$
since $wd_1(U, \{U,D\}, \{L,R\})$ does not hold.


This shows that the optimality notions $wd_{i}$ and $mwd_{i}$ cannot
be justified in the used epistemic framework as `stand alone' concepts
of rationality.


\section{Consequences of Common Knowledge of Rationality}

In this section we show that common knowledge of rationality is
sufficient to entail the customary iterated elimination of strictly
dominated strategies.  We also show that weak dominance is not
amenable to such a treatment.


Given a sequence of properties $\phi := (\phi_1, \LL, \phi_n)$,
we introduce an operator $U_{\phi}$ on the restrictions of
$H$ defined by
\[
U_{\phi}(G) := G',
\]
where $G := (G_1, \LL, G_n)$, $G' := (G'_1, \LL, G'_n)$, and for all $i \in \{1, \LL, n\}$
\[
G'_i := \{ s_i \in G_i \mid \phi_i(s_i, G_i, G_{-i})\}.
\]
So when defining the set of strategies $G'_i$ we use in the second
argument of $\phi_i$ the set $G_i$ of player's $i$ strategies in the
\emph{current} restriction $G$.  That is, $U_{\phi}(G)$
determines the `locally' $\phi$-optimal strategies in $G$.
In contrast, $T_{\phi}(G)$ determines the `globally'
$\phi$-optimal strategies in $G$, in that each player $i$
must consider all of his strategies $s'_i$ that occur in his strategy
set $H_i$ in the \emph{initial game} $H$.


So the `global' form of optimality coincides with rationality, as
introduced in Subsection \ref{subsec:iter}, while the customary
definition of iterated elimination of strictly (or weakly) dominated
strategies refers to the iterations of the appropriate instantiation
of the `local' $U_{\phi}$ operator.






Note that the $U_{\phi}$ operator is non-monotonic for all non-trivial optimality
notions $\phi_i$ such that $\phi_i(s_i, \{s_i\}, (\{s_j\}_{j \neq
  i}))$ for all joint strategies $s$, so in particular for $br_i, sd_i, msd_i, wd_i$ and $mwd_i$.
Indeed, given $s$ let $G_s$ denote the corresponding
restriction in which each player $i$ has a single strategy $s_i$.
Each restriction $G_s$ is a fixpoint of $U_{\phi}$. By non-triviality of $\phi_i$s we have
$U_{\phi}(H) \neq H$, so for each restriction $G_s$ with $s$ including an
eliminated strategy the inclusion $U_{\phi}(G_s) \sse U_{\phi}(H)$
does not hold, even though $G_s \sse H$.  In contrast, as we saw, by
virtue of Lemma \ref{lem:mono} the $T_{\phi}$ operator is monotonic
for $br_i, sd_i$ and $msd_i$.

First we establish the following consequence of
Theorem \ref{thm:epist1}.  When each property $\phi_i$
equals $\textit{br}_{i}$, we write here $\textbf{RAT}({\textit{br}})$
and similarly with $U_{sd}$. 

\newpage

\begin{corollary} \label{thm:just}
\mbox{}

\begin{enumerate} \smallromani
\item For all belief models
\[
G_{\textbf{RAT}({\textit{br}}) \cap B^* \textbf{RAT}({\textit{br}})} \sse U^{\infty}_{\textit{sd}}
\]

\item
for all knowledge models
\[
G_{K^* \textbf{RAT}({\textit{br}})} \sse U^{\infty}_{\textit{sd}}
\]
\end{enumerate}
where in both situations we use in $br_i$ the set of poinr beliefs.
\end{corollary}

\Proof

\NI
$(i)$
By Lemma \ref{lem:mono} and Theorem \ref{thm:epist1}$(i)$
$G_{\textbf{RAT}(\textit{br}) \cap B^* \textbf{RAT}(\textit{br})} \sse T^{\infty}_{\textit{br}}$
Each best response to a joint strategy of the opponents is not
strictly dominated, so for all restrictions $G$
\[
T_{\textit{br}}(G) \sse T_{\textit{sd}}(G)
\]
Also, for all restrictions $G$, $T_{\textit{sd}}(G)
\sse U_{\textit{sd}}(G)$.
So by Lemma \ref{lem:inc} 
$T^{\infty}_{\textit{br}} \sse U^{\infty}_{\textit{sd}}$,
which concludes the proof.
\II

\NI
$(ii)$
By part $(i)$ and the fact that
$K^* \textbf{RAT}({\textit{br}}) \sse \textbf{RAT}({\textit{br}})$.
\HB
\VV

Part $(ii)$ 
formalizes and justifies in the
epistemic framework used here the often used statement:
\begin{quote}
  common knowledge of rationality implies that the players will choose
  only strategies that survive the iterated elimination of strictly
  dominated strategies
\end{quote}
for games with \emph{arbitrary strategy sets} and \emph{transfinite
  iterations} of the elimination process, and where best response means best response to a point belief.

In the case of finite games Theorem \ref{thm:epist1} implies the
following result.  For the case of independent beliefs it is implicitly
stated in \cite{BD87}, explicitly formulated in \cite{Sta94} (see
\cite[ page 181]{BB99}) and proved using Harsanyi type spaces in
\cite{BF08}.

\begin{corollary} \label{thm:just1}
Assume the initial game $H$ is finite.
\begin{enumerate} \smallromani
\item For all belief models for $H$
\[
G_{\textbf{RAT}({\textit{br}}) \cap B^* \textbf{RAT}({\textit{br}})} \sse U^{\infty}_{\textit{msd}},
\]
\item
for all knowledge models for $H$
\[
G_{K^* \textbf{RAT}({\textit{br}})} \sse U^{\infty}_{\textit{msd}},
\]
\end{enumerate}
where in both situations we use in $br_i$ either the set of point beliefs or
the set of independent beliefs or the set of correlated beliefs.
\end{corollary}

\Proof
The argument is analogous as in the previous proof
but relies on a subsidiary result and runs as follows.

\NI
$(i)$
Denote respectively by $brp_i$, $bri_{i}$ and $brc_{i}$ the best response
property w.r.t.~\emph{point}, \emph{independent} and \emph{correlated}
beliefs of the opponents.  Below $\phi$ stands for either $brp$, $bri$ or $brc$.

By Lemma \ref{lem:mono} and Theorem \ref{thm:epist1}
$G_{\textbf{RAT}(\phi) \cap B^* \textbf{RAT}(\phi)} \sse T^{\infty}_{\phi}$.
Further, for all restrictions $G$ we have both
$
T_{\phi}(G) \sse U_{\phi}(G)
$
and
$
U_{\textit{br}}(G) \sse U_{\textit{bri}}(G) \sse U_{\textit{brc}}(G).
$
So by Lemma \ref{lem:inc}
$T^{\infty}_{\phi} \sse U^{\infty}_{\textit{brc}}$.
But by the result of \cite{OR94}, (page 60) (that is a modification of
the original result of \cite{Pea84}), for all restrictions $G$ we have
$U_{\textit{brc}}(G) = U_{\textit{msd}}(G)$, so
$U^{\infty}_{\textit{brc}} = U^{\infty}_{\textit{msd}}$,
which yields the conclusion.
\II

\NI
$(ii)$ By $(i)$ and the fact that
$K^* \textbf{RAT}({\textit{br}}) \sse \textbf{RAT}({\textit{br}})$.
\HB
\VV

Finally, let us clarify the situation for the remaining two
optimality notions, $wd_{i}$ and $mwd_{i}$. For them the
inclusions of Corollaries \ref{thm:just} and \ref{thm:just1} do not hold.
Indeed, it suffices to consider the following initial game $H$:
\begin{center}
\begin{game}{2}{2}
       & $L$    & $R$\\
$U$   &$1,0$   &$1,0$\\
$D$   &$1,0$   &$0,0$
\end{game}
\end{center}

Here every strategy is a best response but $D$ is weakly dominated by $U$. So
both $U^{\infty}_{\textit{wd}}$ and $U^{\infty}_{\textit{mwd}}$
are proper subsets of $T^{\infty}_{\textit{br}}$. On the other hand by
Theorem \ref{thm:epist1}$(iii)$ for some standard knowledge model for $H$ we have
$G_{K^* \textbf{RAT}({\textit{br}})} = T^{\infty}_{\textit{br}}$.
So for this knowledge model neither
$G_{K^* \textbf{RAT}({\textit{br}})} \sse U^{\infty}_{\textit{wd}}$
nor $G_{K^* \textbf{RAT}({\textit{br}})} \sse U^{\infty}_{\textit{mwd}}$ holds.

\section*{Acknowledgements}

We thank one of the referees for useful comments.  We acknowledge
helpful discussions with Adam Brandenburger, who suggested Corollaries
\ref{thm:just} and \ref{thm:just1}, and with Giacomo Bonanno who,
together with a referee of \cite{Apt07}, suggested to incorporate
common beliefs in the analysis.  Joe Halpern pointed us to
\cite{MS89}.  This paper was previously sent for consideration to
another major game theory journal, but ultimately withdrawn because of
different opinions with the referee. We would like to thank the
referee and associate editor of that journal for their comments and
help provided.

\bibliography{/ufs/apt/bib/e,/ufs/apt/bib/apt}
\bibliographystyle{abbrv}
\end{document}